\documentclass[%
 reprint,
 amsmath,amssymb,
 aps,
nofootinbib
]{revtex4-2}

\pdfoutput=1 
\usepackage{graphicx}
\usepackage{dcolumn}
\usepackage{bm}
\usepackage[caption=false]{subfig}
\usepackage{amssymb}
\usepackage{amsmath}
\usepackage{commath}
\usepackage{graphicx,bm}
\usepackage{verbatim}

\usepackage[colorlinks=true,linkcolor=blue,citecolor=blue,allcolors=blue]{hyperref}%

\usepackage{color}
    \definecolor{darkgreen}{rgb}{0,0.5,0}
    \definecolor{darkred}{rgb}{0.5,0,0}
    \definecolor{darkblue}{rgb}{0,0,0.6}
    \definecolor{purple}{rgb}{0.4,.2,0.7}

\newcommand{\bigO}[1]{\ensuremath{\mathcal{O}(#1)}}
\newcommand{\inv}[1]{\frac{1}{#1}}

\newcommand*\diff{\mathop{}\!\mathrm{d}}

\begin{document}

\preprint{APS/123-QED}

\title{Holographic Batteries}

\author{William~D.~Biggs}
\author{Jorge~E.~Santos}%
\affiliation{%
 Department of Applied Mathematics and Theoretical Physics, University of Cambridge, Cambridge, CB3 0WA, United Kingdom
}%

\date{\today}

\begin{abstract}
We study a three-dimensional holographic CFT under the influence of a background electric field on a spacetime containing two black hole horizons. The electric background is fixed such that there is potential difference between the two boundary black holes, inducing a conserved current. By constructing the holographic duals to this set-up, which are solutions to the Einstein-Maxwell equations with a negative cosmological constant in four dimensions, we calculate, to a fully non-linear level, the conductivity of the CFT in this background. Interestingly, we find that the conductivity depends non-trivially on the potential difference. The bulk solutions are flowing geometries containing black hole horizons which are non-Killing and have non-zero expansion. We find a novel property that the past boundary of the future horizon lies deep in the bulk and show this property remains present after small perturbations of the temperature difference of the boundary black holes.
\end{abstract}

\maketitle

\subparagraph{Introduction.}
Given a quantum field theory (QFT), a natural avenue of investigation is to test how it behaves under an external electric field. Studying such behaviour for a strongly coupled QFT using direct field theory techniques is computationally very challenging (though progress has been made when the system is in the proximity of quantum critical points \cite{green.95.267001}; see also \cite{PhysRevLett.109.091601} for a holographic description of a similar setup). However, since its advent, the AdS/CFT correspondence \cite{Maldacena:1997re,Witten:1998qj,Aharony:1999ti} has allowed for the indirect study of strongly coupled condensed matter systems via gravitational calculations (see for instance \cite{Hartnoll:2018xxg} for an excellent review on the topic).

Specifically, we work in the limit of the AdS/CFT duality in which the gravitational theory is well described by classical gravity. In this limit, the duality maps a problem of studying a strongly coupled CFT with a large number of degrees of freedom living on a fixed (but possibly curved) background, $\mathcal{B}$, to a gravitational problem in which one must find a corresponding \textit{asymptotically locally AdS}, (AlAdS), spacetime, called the \textit{bulk}, which possesses a conformal boundary on which the induced metric is conformal to $\mathcal{B}$ (see Ref. \cite{Marolf:2013ioa} for an introduction to this method). 

Taking $\mathcal{B}$ to be a black hole background has allowed for the study of Hawking radiation at strong coupling. The dual bulk solutions are generally called \textit{droplets} and \textit{funnels} \cite{Hubeny:2009ru, Hubeny:2009kz,Hubeny:2009rc,Caldarelli:2011wa, Figueras:2011va,Santos:2012he, Santos:2014yja,Fischetti:2016oyo, Santos:2020kmq,Fischetti:2012ps,Fischetti:2012vt,Marolf:2019wkz,Emparan:2021ewh,Biggs:2022lvi}, with the distinction between these two classes originating from the structure of the horizons in the bulk. A bulk solution with a horizon connecting two distinct boundary horizons is called a funnel, whereas a bulk solution with horizons each emanating from only a single boundary horizon is called a droplet.

\begin{figure}[b!]
    \centering
    \includegraphics[width=0.49\textwidth]{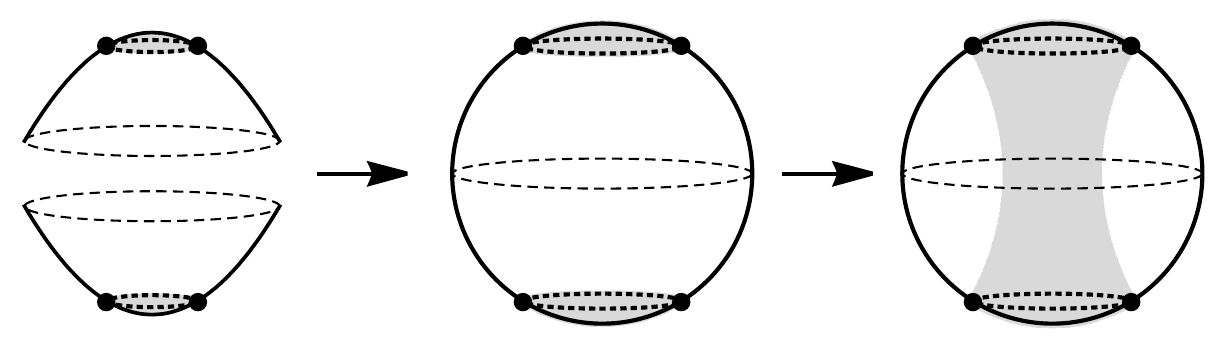}
    \caption{Some sketches of spatial cross-sections of spacetimes of interest. \textbf{Left:} Two BTZ black holes with the dashed curves being their asymptomatic boundaries and the dotted line their horizons. The interior is shaded. \textbf{Middle:} The boundary geometry found by patching the BTZ spacetimes together at infinity and then compactifying onto the ESU. \textbf{Right: } The global funnel, a solution in the bulk with a horizon connecting the two boundary horizons.}
    \label{fig:funnel_sketch}
\end{figure}

In this letter, we will focus on four-dimensional global funnels. The boundary geometry will be given by the conformal compactification of the geometry obtained by ``patching together" two identical Bañados-Teitelboim-Zanelli (BTZ) black holes (shown on the left in Fig. \ref{fig:funnel_sketch}) at infinity. This yields two black holes antipodally situated in the Einstein Static Universe (ESU) as sketched in the middle in Fig. \ref{fig:funnel_sketch}. There is a well-known solution called the BTZ black string or uniform funnel which connects these two boundary black holes, however, even in the case of vacuum gravity in the bulk, there is a rich structure of bulk solutions 
 beyond this uniform funnel \cite{Fischetti:2012ps, Fischetti:2012vt, Marolf:2019wkz, Emparan:2021ewh}. The right hand sketch in Fig. \ref{fig:funnel_sketch} is a schematic drawing of a global funnel.

We add a chemical potential to the field theory which induces a conserved current, $J^\mu$, and, on the gravitational side of the duality, causes a deformation of the bulk geometry away from the uniform funnel to a new solution to the Einstein-Maxwell equations with a negative cosmological constant.
We will define the chemical potential to vary on the boundary background, and in particular fix that it approaches two different values at the two BTZ black hole horizons in the boundary geometry. Since these charged global funnels correspond to the QFT sourced by the two horizons at the same temperature but with a potential difference between them, we dub these solutions as \textit{holographic batteries}.

Calculating the bulk solution allows one to extract the conserved current, $J^\mu$, induced by the source on the field theory side, and so this provides a process to calculate the non-linear conductivity of the boundary field theory, going beyond the linear regime examined in \cite{Horowitz:2012ky,Horowitz:2012gs,Blake:2013owa,Andrade:2013gsa,Donos:2013eha,Donos:2014cya,Donos:2014yya,Rangamani:2015hka,Davison:2015bea}. Interestingly, we find that the conductivity is not a constant value, \textit{i.e.} the current, $J^\mu$, depends non-trivially on the magnitude of the chemical potential, despite there being no net current, in contrast to what was observed in \cite{Horowitz:2013mia,Withers:2016lft}.

Moreover, the chemical potential varying across the boundary geometry induces classical flow along the bulk horizon. Unlike previously found solutions containing flowing horizons \cite{Figueras:2012rb, Fischetti:2012vt, Santos:2020kmq, Biggs:2022lvi}, this flow is not caused by a temperature difference between two asymptotic regions of the bulk horizon, and this means the properties of the holographic batteries are subtly different to these other flowing solutions. In particular, we show that the past boundary of the bulk horizon lies deep in the bulk, as opposed to on one of the points at which the bulk horizon is anchored on the boundary, as seen in all previous flowing black hole geometries. We show that this property is generic by considering holographic batteries in which the boundary black holes can also have a small temperature difference.
\subparagraph{Finding the batteries.}
First let us consider the metric of the BTZ black hole:
\begin{equation}
    \diff s^2_{BTZ} = -f(r)\diff T^2 + \frac{\diff r^2}{f(r)}+r^2 \diff\varphi^2,
\end{equation}
with $f(r)=(r^2 - r_0^2)/\ell_3^2$, where $r_0$ is the radius of the BTZ black hole and $\ell_3$ is the the three-dimensional AdS length scale. Taking
\begin{equation}\label{eq:newcoord}
    r = \frac{r_0}{x\sqrt{2-x^2}},\quad T = \frac{\ell_3^2}{r_0}\,t, \quad \varphi = \frac{\ell_3}{r_0}\phi,
\end{equation}
so that $x=1$ is the horizon and $x=0$ is infinity, yields 
\begin{align}\label{eqn:bndy_metric}
    \diff \tilde{s}^2_{BTZ} &= \Omega(x)^2 \diff s^2_{BTZ} \nonumber\\ =\,&\ell_3^2\left[-\diff t^2 +\inv{\left(1-x^2\right)^2}\left(\frac{4 \diff x^2}{2-x^2} +\diff \phi^2\right)\right],
\end{align}
where we've multiplied the metric by a conformal factor
\begin{equation}
    \Omega(x) = \frac{x\sqrt{2-x^2}}{1-x^2}.
\end{equation}
Note that $\phi$ has a period of $2 \pi r_0 / \ell_3$, which we must take into account when calculating any global properties of the solutions. The geometry upon which we wish to study the CFT is obtained by patching two copies of the metric given by $\diff \tilde{s}^2_{BTZ}$ at their boundaries, \textit{i.e.} at $x=0$. Thus the boundary metric is given by the metric in (\ref{eqn:bndy_metric}), with $x\in[-1,1]$ and $x=\pm 1$ being the two boundary black hole horizons. The temperature of both horizons, measured in units of the original $T$ coordinate, is given by $T_H = r_0 / (2\pi \ell_3^2)$.

The idea is to add an electric field on this background which acts as a source and to fix the chemical potential at the two boundary horizons, $x = \pm 1$. Specifically, let us add an electrical source given by the following vector potential:
\begin{equation}\label{eqn:source}
    A^{(0)} = \mu\, g(x) \diff t,
\end{equation}
where $g(x)$ is a profile we are free to choose and which we will design so that $g(1)=+1$ and $g(-1)=-1$. The potential difference between the two horizons is
\begin{equation}
    V :=  \left[A^{(0)}\cdot k \right]^{x=1}_{x=-1} = 4 \pi \mu T_H ,
\end{equation}
where $k = \partial / \partial T$. 
For the majority of this letter we take
\begin{equation}\label{eq:chemical}
    g(x):= \sin\left(\frac{\pi}{2}x\sqrt{2-x^2}\right),
\end{equation}
which we call the \textit{sine profile}.
\paragraph*{The Ansatz.}
The addition of an electric source in the boundary theory means that the bulk theory is Einstein-Maxwell with a negative cosmological constant in four dimensions, with the following equations of motion:
\begin{subequations}\label{eqn:eom}
\begin{align} \label{eqn:Einstein}
     0 &= E_{ab}:=R_{ab} + \frac{3}{\ell_4^2}g_{ab} - 2 T_{ab} \\
     0 &=\nabla^a F_{ab} \label{eqn:Maxwell}
\end{align}
\end{subequations}
where $\ell_4$ is the four-dimensional AdS length scale, $F = \diff A$ is the field strength tensor of the Maxwell field and the bulk stress tensor is given by 
\begin{equation}
    T_{ab} = F_a{}^cF_{bc}-\inv{4}g_{ab}F^{cd}F_{cd}.
\end{equation}
We begin with an \textit{Ansatz} in Bondi-Sachs gauge which possesses a null hypersurface at $y=1$ and a conformal boundary at $y=0$. We also assume the solutions will be stationary and axisymmetric, with corresponding Killing vector fields $\partial_v$ and $\partial_\phi$, respectively. For the gauge field we pick a gauge in which $A_y = 0$. In such a gauge, the \textit{Ansatz} is given by
\begin{subequations}\label{eq:ansatz}
    \begin{align}\label{eq:metric_ansatz}
    \diff s^2& = \frac{\ell_4^2}{y^2}\left[ \rule{0cm}{0.7cm}
    q_2^2\left(-(1-y^2)q_1 \diff v^2 - 2 \diff v \diff y \right) \right. \nonumber  \\
    &\left. +\frac{q_5^2}{\left(1-x^2\right)^2}\left(\frac{4\left(\diff x -(1-x^2)q_4 \diff v\right)^2}{(2-x^2)q_3} + q_3 \diff \phi^2\right)\right] 
\end{align}
and
\vspace{-10pt}
\begin{equation}\label{eq:potential_ansatz}
    A = \ell_4 \left( q_6 \diff v + \frac{q_7}{1-x^2}\diff x\right),
\end{equation}
\end{subequations}
where $q_i(x,y)$ are unknown functions which depend upon $x$ and $y$. Schematically, this spacetime also looks like the right-hand sketch in Fig. \ref{fig:funnel_sketch}.

There is still some gauge freedom in the \textit{Ansatz} which can be used to fix the radial dependence of either $q_2$ or $q_5$ completely. In our case we fix that of $q_5$, by enforcing
\begin{equation}\label{eqn:q5}
q_5(x,y) = 1+y^2 S_2(x).
\end{equation}
With such an \textit{Ansatz}, the partial differential equations arising from the equations of motion can be solved numerically as a boundary value problem after setting suitable boundary conditions, as was first set out, and more fully explained, in \cite{Biggs:2022lvi} for the case of pure gravity and \cite{Horowitz:2022mly} for Einstein-Maxwell. We have briefly summarised the integration scheme in the supplementary material. \cite{Israel:1964a} 
Let us emphasise here that at the conformal boundary, $y=0$, we set Dirichlet boundary conditions such that the induced metric is conformal to the metric given by (\ref{eqn:bndy_metric}) and the bulk vector potential is equal, up to a factor of $\ell_4$, to the boundary source, (\ref{eqn:source}).

We also obtained the solutions using the DeTurck method \cite{Headrick:2009pv, Wiseman:2011by, Dias:2015nua}, and present the \textit{Ansatz} for the solutions in that gauge in the supplementary material. 
\subparagraph{Results.}
The holographic stress tensor, $\langle T_{\mu\nu} \rangle$, and conserved current, $\langle J^\mu\rangle$, can be extracted from the numerical solutions using the standard procedure of holographic renormalization \cite{deHaro:2000vlm}, which we describe explicitly in the supplementary material. One benefit of the Bondi-Sachs gauge over the DeTurck gauge is that no non-analyticities arise in this procedure.
\paragraph*{Conductivity.}
Of particular interest will be the current,  $\langle J^\mu \rangle$, which is conserved:
\begin{equation}\label{eqn:Jcons}
    \mathrm{D}_\mu \langle J^\mu \rangle = 0,
\end{equation}
where $\mathrm{D}$ is the covariant derivative associated to the boundary metric, given by (\ref{eqn:bndy_metric}). We find that
\begin{equation}
    \langle J_x \rangle 
    = \frac{\nu\,C_1}{\ell_3\sqrt{2-x^2}},
\end{equation}
where $\nu$ is a dimensionless quantity which depends upon the number of degrees of freedom of the CFT and is defined holographically in terms of the dual parameters of the gravitational theory by
\begin{equation}
    \nu = \frac{\ell_4^2}{4\pi G_4},
\end{equation}
with $G_4$ being Newton's gravitational constant for the four-dimensional theory. Moreover, as is explained explicitly in the supplementary material, $C_1$ is an integration constant that arises from a local analysis of the equations of motion near the conformal boundary.
Specifically, one can show that 
\begin{equation}\label{eq:dyq7}
    \partial_y q_7(x,0) = (1-x^2)\left(\frac{C_1}{\sqrt{2-x^2}} - \mu \,g'(x)\right).
\end{equation}
Though the local analysis of the equations of motion fix that $\partial_y q_7$ must take this functional form near the boundary, 
one must solve the equations fully, after enforcing regularity deep in the bulk, in order to extract the precise value of $C_1$.

We can define the \textit{total current}, $I$, by integrating $\langle J^\mu \rangle$ over a circle, $S^1_x$, of fixed $x$ at a fixed time-slice in the boundary geometry: 
\begin{align}
    I :&= \int_{S^1_x} \diff \phi \sqrt{g_{\phi\phi}}\,m_\mu \langle J^\mu \rangle \nonumber \\
    &= 2\nu \pi^2 T_H C_1
\end{align}
where $m^\mu$ unit normal is the unit normal to the circle $S^1_x$. Let us note again that $\phi$ has periodicity $2 \pi r_0 / \ell_3$, which must be taken into account when computing this integral. Note that the value of $I$ is independent of the choice of $x$ at which one fixes the circle, $S^1_x$, which follows as a direct consequence of (\ref{eqn:Jcons}).

\begin{figure}[t!]
    \centering
    \includegraphics[width=0.49\textwidth]{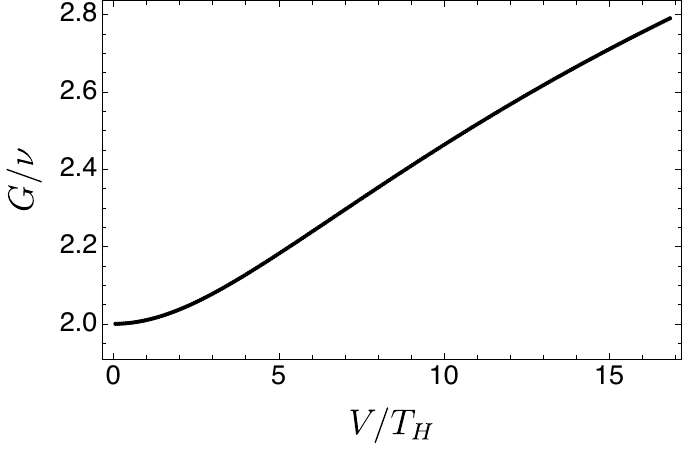}
    \caption{The conductance of the holographic batteries with the sine profile. Each black dot is a different solution with a different value of the parameter $\mu$. We plot the conductance, $G$, against the potential difference, $V$, normalised by the temperature of the black hole horizons, $T_H$. 
    }
    \label{fig:conductance}
\end{figure}

We can describe the \textit{conductance}, $G$, of the holographic battery by dividing the total current by the potential difference between the two horizons, \textit{i.e.}
\begin{equation}\label{eqn:conductance}
    G := \frac{I}{V} = \frac{\nu \pi C_1}{2\mu}.
\end{equation}
The conductance depends upon both the choice of profile, $g(x)$, and the magnitude of the chemical potential, $\mu$, or equivalently the potential difference between the two horizons, $V$. In Fig. \ref{fig:conductance}, we plot the conductance of the holographic batteries, with the sine profile defined by (\ref{eq:chemical}), against the potential difference.
As $V/T_H \to 0$, the conductance tends to two and it increases with the voltage. Moreover, the gradient of the curve at $V = 0$ is zero, meaning that one has to go beyond the linear regime in order to see the non-trivial dependence of the conductance on the potential difference. The derivative of the conductance with respect to $V/T_H$ possesses a turning point, in this case at $V/T_H \simeq 6.62$, with the gradient of the curve decreasing for larger values of $V/T_H$. It would be of interest to investigate further what happens to $G$ as $V/T_H$ becomes very large, though there are numerical challenges in extending to this region of the parameter space.

The behaviour of the conductance appears qualitatively similar for other choices of the profile, $g(x)$, which each satisfy $g(1)=1=-g(-1)$. 
In the supplementary material we provide plots of the conductance for other such profiles as well as a proof that $G/\nu \to 2$ as $V/T_H \to 0$ for \textit{any} choice of odd profile $g(x)$ by conducting an analysis of the linearised equations of motion.

One can also compute the local \textit{conductivity} of the field by computing the ratio between the $x$-components of the induced current and the source electric field,
\begin{align}
    \sigma(x):=&\frac{\langle J^x \rangle}{(F^{(0)})^{tx}} \nonumber \\
    =& \frac{2 G}{\pi} \cdot \inv{ g'(x)\sqrt{2-x^2}}.
\end{align}
Since the external electric field is not homogeneous, the conductivity is not constant, however, for a given profile its functional form is always the same, with its magnitude determined by the value of the overall conductance, $G$.
\paragraph*{Energy Flow.}
The addition of a chemical potential causes some heating of the dual CFT by the Joule effect and hence there is flow in the boundary field theory. As was observed  in other flowing solutions \cite{Figueras:2012rb, Fischetti:2012vt, Santos:2020kmq, Biggs:2022lvi}, this means that the bulk horizon is not a Killing horizon. The flow can be expressed via an integral of the holographic stress tensor over a circle, $S^1_x$, of fixed $x$ in the boundary geometry as follows:
\begin{align}\label{eq:flow}
    \Phi(x)&=-\int_{S^1_x}\diff \phi \sqrt{-\gamma}\,m_\mu k^\nu \langle T^{\mu}{}_{\nu} \rangle \nonumber\\
    &= 2 \pi T_H \left( 2 G\mu^2 g(x) +3 \pi \nu C_2 \right),
\end{align}
where $\gamma_{\mu\nu}$ is the induced metric on a constant $x$ slice of the boundary geometry with determinant $\gamma$ and unit normal $m_\mu \propto (\diff x)_\mu$, whilst $k^\mu \propto (\diff t)^\mu$ is the normalised stationary Killing vector field. Note, therefore, that the flow is simply proportional to the $T^x{}_t$ component of the holographic stress tensor, which is given in the supplementary material. The constant $C_2$ is a coefficient in the asymptotic expansion of the $q_4$ function, which is not fixed by a local analysis of the equations of motion.

In the current case, in which the two boundary horizons have the same temperature and the chemical potential, $g(x)$, is odd, we find empirically, as one would expect, that $C_2 = 0$, and hence there is no net flow between the two horizons. Thus, the flow, $\Phi(x)$, is proportional to the chemical potential, and is odd with $\Phi(x) >0 $ for $x>0$, meaning that there is flow in both directions originating from the point $x=0$ and moving outwards towards the boundary. As we will see this behaviour near $x=0$ has an interesting effect on the structure of the horizon of the bulk geometry.
\paragraph*{Properties of the bulk horizon.}
By design, the $y=1$ is a null hypersurface. We checked explicitly (in a similar manner to in \cite{Biggs:2022lvi}) that there exist future-directed radial null curves from anywhere outside of this hypersurface to the boundary, suggesting it is the event horizon. Let us consider a generator of the horizon, $U_a \propto (\diff y)_a$, which can be parameterised by the $x$ coordinate. The affine parameter, $\lambda(x)$, can be obtained from the geodesic equation or from Raychaudhuri's equation.

At the axis of symmetry at $x=0$, we find that $\lambda'(x) \to 0$, with $\lambda(0)$ taking a finite value, which we are free to choose via an affine transformation as $\lambda(0)=0$. Moreover, we find that $U^a$ is future-directed in both directions moving away from $x=0$. This suggests that the horizon is better thought of as being generated by two separate future-directed generators, both described by $U^a$ and originating at $x=0$, one moving in the positive $x$-direction and the other moving in the negative $x$-direction. Hence, the point $x=0$ is the \textit{past boundary of the future horizon} of the solutions.

\begin{figure}[t!]
    \centering
    \includegraphics[width=0.49\textwidth]{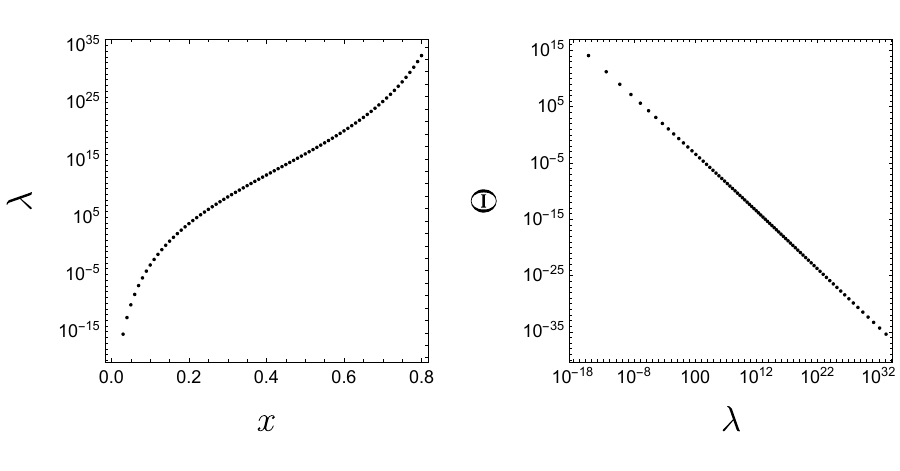}
    \vspace{-20pt}
    \caption{\textbf{Left:} The affine parameter, $\lambda$, along the generator of the $x > 0$ region of the horizon, plotted in a log plot against the coordinate $x$, for a holographic battery with $\mu = 0.24$. Here we have made the choice that $\lambda(0)=0$ and $\lambda'(0.1)=0.1$. \textbf{Right:} The expansion, $\Theta$, along the generator for this holographic battery, plotted in a log-log plot against $\lambda$. The expansion diverges at $\lambda = 0$.}
    \vspace{-10pt}
    \label{fig:affine_and_expansion}
   
\end{figure}

Let us restrict to the $x\geq 0$ region of the horizon, since the behaviour in the $x \leq 0$ region is similar by symmetry. In the left-hand panel of Fig. \ref{fig:affine_and_expansion}, we plot the affine parameter along the generator against the $x$ coordinate in a log plot.

Given our affinely-parameterised null geodesic $U^a$, the $B$-tensor is given by $B_{IJ}=\nabla_I U_J$, where $I$ and $J$ run over $\{v,\phi\}$. The expansion, $\Theta$, and shear, $\sigma_{IJ}$, are the trace and symmetric-traceless parts of this tensor, respectively.
The fact that $\lambda'(x)\to 0$ as $x \to 0$ means that the expansion and shear diverge as $x \to 0$ along the horizon. In the right-hand panel of Fig. \ref{fig:affine_and_expansion} we have plotted the expansion against the affine parameter along the generator of the $x>0$ region of the horizon.

In flowing horizons, the past boundary is generally situated at the point at which the flow along the horizon emanates from and, moreover, the expansion diverges at this point. In the previous cases \cite{Fischetti:2012vt, Santos:2020kmq, Biggs:2022lvi}, the flow was induced by a temperature difference between the boundary black holes meaning that the past boundary tended to be the hotter boundary horizon. However, in the current case there is no temperature difference; the flow is instead induced by the Joule effect, emerging from $x=0$ and proceeding outwards, towards the boundary black holes, situated at $x =\pm 1$. Hence, our case is distinct in that the past boundary is situated spatially in the \textit{centre} of the horizon with the future-directed horizon generators extending outwards in either direction. This has another interesting consequence: at $x=0$ (and only at $x=0$), the generator $U^a$ coincides with the stationary Killing vector, $U^a \propto (\partial_v)^a$, meaning that the Killing vector is a generator of the horizon only on a proper submanifold of the horizon. To our knowledge, this feature has so far not been found to occur in any other instances of flowing horizons.

The divergent expansion at $x=0$ suggests that the tidal forces between neighbouring horizon generators diverges at this point. However, since the future-directed generators emerge from this point, this singularity is always in their far past. If we instead consider the geodesics of infalling observers, then we find no infinite tidal forces are felt for such neighbouring geodesics, even near $x=0$, hence these solutions do not contain any physical singularities.
\paragraph*{Detuning the temperatures.}
One may wonder whether this property of the past boundary of the future horizon of the holographic batteries lying deep in the bulk is generic or simply a product of the symmetry of the set-up. To investigate this, we detune the temperatures by adding a non-trivial profile to the $g_{tt}$ component of the boundary geometry:
\begin{subequations}
    \begin{align}\label{eqn:detuned_bndy_metric}
        \diff s ^2_{detuned} =&- \ell_3^2 h(x)^2\diff t^2 + \frac{\ell_3^2}{\left(1-x^2\right)^2}\left(\frac{4 \diff x^2}{2-x^2} + \diff \phi^2 \right)
    \end{align}
    where
    \vspace{-10pt}
    \begin{equation}
        h(x) := 1 + \beta \sin \left(\frac{\pi}{2}x \sqrt{2-x^2}\right),
    \end{equation}
\end{subequations}
so that the ratio between the temperatures of the two boundary horizons is $(1+\beta) / (1-\beta)$. The \textit{detuned holographic batteries} are the bulk duals to the CFT on this background, still under the influence of an additional chemical potential, given by (\ref{eq:chemical}). The method to find the detuned solutions is almost identical to the previous, tuned case, with the \textit{Ansatz} only slightly modified to accommodate for the non-trivial $h(x)$ profile. Note, one can recover the tuned holographic batteries by taking $\beta = 0$.

\begin{figure}[t!]
    \centering
    \includegraphics[width=0.47\textwidth]{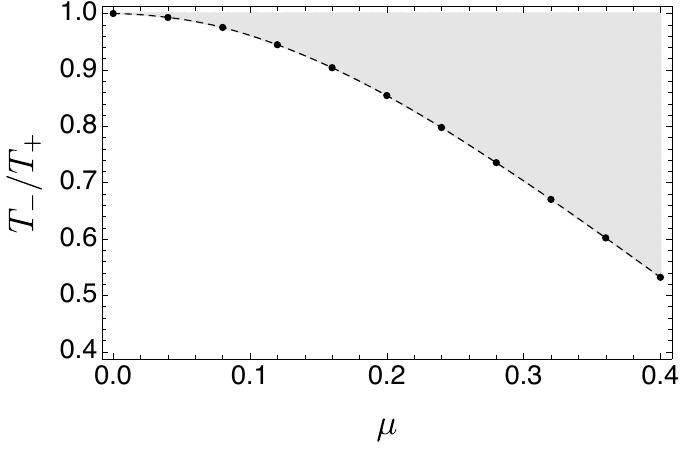}
    \caption{The region in the parameter space of the detuned holographic batteries in which the past boundary of the horizon lies in the bulk rather than being anchored on the conformal boundary at the hotter boundary black hole. The chemical potential is fixed to be $\pm \mu$ on the horizon with temperature $T_\pm$, respectively, where without loss of generality $T_+>T_-$. 
    }
    \label{fig:Tmu}
\end{figure}

Interestingly, we find that there is an open set in the parameter space, depicted by the shaded region in Fig. \ref{fig:Tmu}, in which the flow vanishes at some point along the horizon and hence the past boundary of the future horizon lies deep in the bulk. Roughly speaking, for these solutions, the contribution to the flow from the temperature difference is not large enough to everywhere overcome the flow due to the Joule effect and cause the flow to be in the same direction throughout the boundary geometry.

\paragraph*{Discussion of dominance.} Finally, one may wonder whether the charged funnels dominate over a solution with the same boundary geometry in which two droplets emerge from the boundary horizons but do not connect in the bulk, though it is yet to be explicitly shown that such a solution exists for $\mu \neq 0$. Due to the fact the funnels are flowing solutions, there will be difficulty in using thermodynamic arguments since the free energy is not well-defined (a problem also faced in \cite{Biggs:2022lvi}). However, it is known that the uniform funnel dominates for a BTZ black hole with a large enough radius \cite{Hubeny:2009rc}, and physically one would expect that by adding a charge difference between the two boundary black holes one would remain in the phase in which the funnel dominates.

\subparagraph*{Acknowledgments:}
We would like to thank Sean Hartnoll, Gary Horowitz and Don Marolf for reading an earlier version of this paper and for providing critical comments. We would also like to thank Sean Hartnoll for drawing our attention to Ref. \cite{green.95.267001}. W.~B. was supported by an STFC studentship, ST/S505298/1, and a Vice Chancellor's award from the University of Cambridge, and J. E. S. has been partially supported by STFC consolidated grant ST/T000694/1. The numerical component of this study was carried out using the computational facilities of the Fawcett High Performance Computing system at the Faculty of Mathematics, University of Cambridge, funded by STFC consolidated grants ST/P000681/1, ST/T000694/1 and ST/P000673/1.
\newpage
\bibliographystyle{utphys-modified}
\bibliography{papers}

\clearpage
\onecolumngrid
\appendix

\section*{Supplemental material}

\section{The Integration Scheme in Bondi-Sachs Gauge}

The Einstein-Maxwell equations yield ten equations for the seven unknown functions in the metric and vector potential given by the \textit{Ansatz}, (\ref{eq:ansatz}), in Bondi-Sachs gauge. On the face of it, one may worry that the equations are over-determined, but in reality not all of the equations of motion are independent. In fact, one can solve the \textit{bulk equations}, which are given by $E_{ij} = 0$ and $\nabla^a F_{ai}=0$ for $i,\, j \neq v$ where $v$ is the temporal direction, throughout the whole of the space and the remaining equations, named the \textit{supplementary equations}, $E_{va} = 0$ and $\nabla^a F_{av}=0$, on a constant $y$ slice, which in our case we choose to be the $y=1$ null hypersurface, and which we check \textit{a posteriori} is the event horizon.

Having set the supplementary equations as Robin boundary conditions at $y=1$, the contracted Bianchi identity, together with the bulk equations, enforce that the supplementary equations must be satisfied throughout the whole spacetime (see, for instance, \cite{Biggs:2022lvi}). It still remains to be fully understood whether this integration scheme leads to a well-defined elliptic PDE problem for a stationary spacetime.

Now let us address the boundary conditions at the remaining boundaries of the integration domain. Recall that the radial dependence of the $q_5$ function is fixed as (\ref{eqn:q5}) as a gauge choice, so we need not prescribe boundary conditions for this function. At $y=0$, the conformal boundary, we set Dirichlet boundary conditions enforcing that the induced metric on the boundary is conformal to (\ref{eqn:bndy_metric}). To do so, we set $q_1 = q_2 = q_3 = 1$ and $q_4 = 0$. Then after taking the transformations:
\begin{equation}
    \diff v = \diff t - \inv{1-y^2}\diff y, \quad y = \frac{z}{\ell_3}
\end{equation}
we find that to leading order in $z$,
\begin{equation}
    \diff s^2 = \frac{L^2}{z^2}\left(\diff z^2 + \diff \tilde{s}_{BTZ}\right) + \bigO{z^{-1}},
\end{equation}
where $\diff \tilde{s}_{BTZ}$ is the desired boundary metric, given by (\ref{eqn:bndy_metric}). As for the gauge field, we want to it to approach the desired electric source (\ref{eqn:source}) at leading order at the conformal boundary. As such, we pick at $y=0$ that
\begin{equation}
    q_6 = \mu\, g(x), \quad q_7 = 0.
\end{equation}

The funnels naturally lie on a coordinate domain with only two boundaries, the conformal boundary and the bulk horizon. However, running numerics on such a domain is very difficult, and so, similarly to what was done in \cite{Fischetti:2012vt, Santos:2020kmq,Biggs:2022lvi}, the two points at which the horizon meets the boundary are ``blown up" to two extra sides of the coordinate domain, given by $x = \pm 1$. This is possible because the horizon must approach the geometry of a hyperbolic black hole asymptotically as it approaches the conformal boundary. Hence at $x = \pm 1$, we fix that the horizon has this limiting behaviour. This is also done by setting $q_1 = q_2 = q_3 = 1$ and $q_4 = 0$ at these boundaries. Moreover, to match the vector potential at these boundaries to the conformal boundary, we take $q_6(1,y) = \mu$, $q_6(-1,y) = -\mu$ and $q_7(\pm1,y) = 0$.

Finally, in order to numerically solve the bulk equations, we use pseudo-spectral collocation methods on a Chebyshev-Gauss-Lobatto grid to approximate the PDEs with non-linear algebraic equations which are solved iteratively by Newton's method (see \cite{Dias:2015nua} for a detailed review of these methods in the context of the Einstein equation).

Due to the fact the supplementary equations are not enforced explicitly throughout the bulk, they give us a way to test the veracity of the solutions, and provide a convergence test since they should go to zero across the bulk in the continuum limit. In Fig. \ref{fig:BS_err}, we plot the maximum value of the supplementary equations across the bulk against the number of lattice points used in the numerics in a log-log plot. The roughly linear behaviour of the convergence in the log-log plot suggests power-law convergence. 

\begin{figure}[t]
    \centering
    \includegraphics[width =0.75\textwidth]{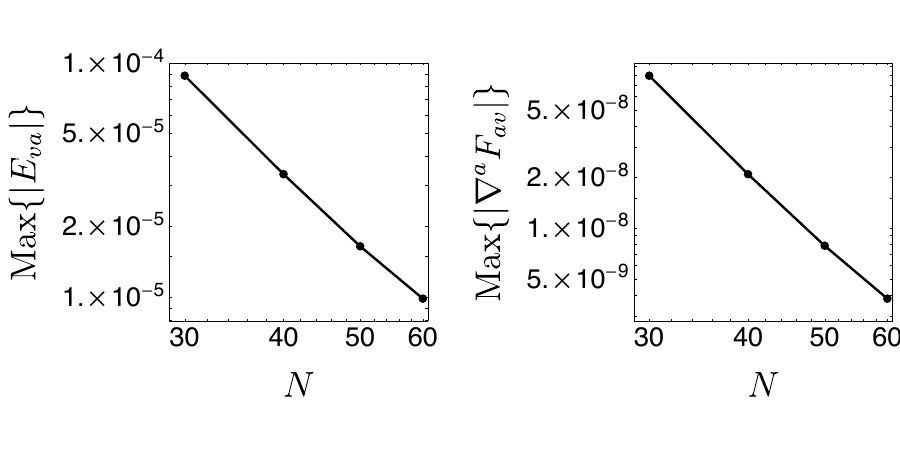}
    \caption{The maximum value of the supplementary equations, $E_{va}$ on the left and $\nabla^a F_{ab}$ on the right, for a holographic battery with $\mu = 0.36$ obtained numerically on a discrete lattice of size $2N \times N$ for different values of $N$, plotted in a log-log plot.}
    \vspace{-8pt}
    \label{fig:BS_err}
\end{figure}

\paragraph*{Holographic renormalization.}
Having set the Dirichlet boundary conditions at the conformal boundary, $y=0$, a local, order-by-order analysis of the equations of motion near $y=0$ determine the asymptotic behaviour of the $q_i$ functions to a large extent. Expanding about this boundary, one discovers that
\begin{subequations}\label{eq:bndy_expansion}
    
\begin{alignat}{4}
    q_1(x,y)&=1 + 3 S_2(x) y^2 + \alpha_1(x)y^3+ \bigO{y^4} \\
    q_2(x,y)&=1-\inv{2}S_2(x)y^2 + \bigO{y^4} \\
    q_3(x,y)&=1+ \alpha_3(x)y^3 + \bigO{y^4} \\
    q_4(x,y)&= \alpha_4(x)y^3 + \bigO{y^4} \\
    q_5(x,y)&=1+y^2 S_2(x) \\
    q_6(x,y) &= \mu\, g(x) + \beta_6(x)y +\bigO{y^2} \\
    q_7(x,y) &= \beta_7(x)y +\bigO{y^2}
\end{alignat}
\end{subequations}
where, we recall that the radial profile of the $q_5$ was actually a gauge choice we were free to make. All higher order terms are fixed in terms of $\{\alpha_1,\,\alpha_3,\,\alpha_4,\,\beta_6,\,\beta_7\}$, which are the free functions that are not fixed by any local analysis of the equations of motion near the boundary. Instead, in order to find these five functions, we need to solve the equations of motion fully, whilst enforcing regularity deep in the bulk. The local analysis of the equations of motion does, however, provide some information on the functional form of $\beta_7$ and $\alpha_4$:
\begin{align}\label{eqn:beta}
    \beta_7(x) &= (1-x^2)\left(\frac{C_1}{\sqrt{2-x^2}} - \mu \,g'(x)\right) \\
    \label{eqn:alpha}
    \alpha_4(x) &= \left(1-x^2\right) 
   \left[\sqrt{2-x^2}\left(C_2 + \inv{3} \mu \,C_1\, g(x)\right) -\inv{4}(2-x^2) S_2'(x)\right],
\end{align}
where $C_1$ and $C_2$ are unknown constants not fixed by the asymptotic expansion, which are intimately connected to the conductivity and the flow of the field theory. Once the $\{\alpha_1,\,\alpha_3,\,\alpha_4,\,\beta_6,\,\beta_7\}$ functions are obtained via the bulk calculation, one can proceed with the standard approach of holographic renormalization \cite{deHaro:2000vlm} by transforming to Fefferman-Graham gauge, in which the metric and vector potential in four dimensions take the form:
\begin{subequations}\label{eq:FG}
\begin{equation}
    \diff s^2=\frac{\ell_4^2}{z^2}\left[\diff z^2+ \left(g^{(0)}_{\mu\nu}+g^{(2)}_{\mu\nu} z^2+ g^{(3)}_{\mu\nu}z^3+ \cdots\right)\diff x^\mu \diff x^\nu\right],
\end{equation}
\begin{equation}
    A = \ell_4\left(A^{(0)}_\mu + zA^{(1)}_\mu \right) \diff x^\mu+\bigO{z^2},
\end{equation}
\end{subequations}
where $g^{(0)}_{\mu\nu}$ is the boundary metric, $A^{(0)}_\mu$ is the boundary electric source and the Greek indices run over all coordinates other than the AdS radial direction, $z$. This can be achieved with a coordinate transformation 
\begin{subequations}\label{eq:FG_transformation}
\begin{align}
    x &= w+\sum_{j=1}^4 \gamma_j(w) z^j, \quad \quad
    y = \sum_{j=1}^4 \delta_j(w) z^j,
\end{align}
as well as a gauge transformation of the vector potential, $A \to A + \diff \chi$, with
\begin{align}
    \chi = \sum_{j=1}^2 \varepsilon_j(w)z^j.
\end{align}
\end{subequations}
The explicit expressions for $\gamma_j(w),\,\delta_j(w)$ and $\varepsilon_j(w)$ can be found by taking the expansion of the metric and gauge field given by (\ref{eq:bndy_expansion}), transforming with (\ref{eq:FG_transformation}) and then matching with the Fefferman-Graham form of the metric and vector potential order-by-order in $z$.

\begin{figure}
    \centering
    \includegraphics[width =0.75\textwidth]{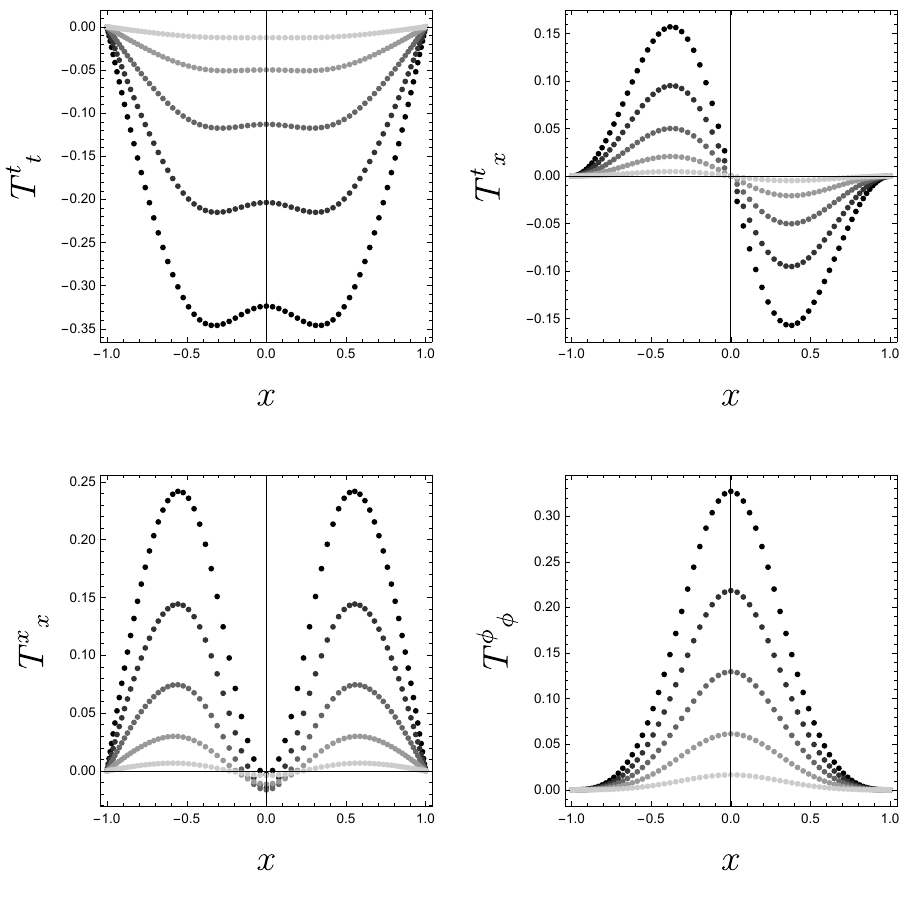}
    \caption{Plots of the holographic stress tensor against the $x$ coordinate. The different curves  correspond to the sine profile for $\mu \in \{0.5,\,1,\,1.5,\,2,\,2.5\}$, with the darker curves corresponding to larger values of $\mu$. The magnitude of each component $T^\mu{}_\nu$ increases as $\mu$ increases.}
    \label{fig:Tmunu}
\end{figure}

Finally, the vacuum expectation value of the holographic stress tensor and the conserved current can be simply read off as
\begin{align}
    \langle T_{\mu\nu} \rangle = \frac{3 \ell_4^2}{16 \pi G_4} g^{(3)}_{\mu\nu}, \quad\quad
    \langle J_\mu \rangle = \frac{\ell_4^2}{4 \pi G_4} A^{(1)}_\mu.
\end{align}
Note that the holographic stress tensor, $\langle T_{\mu\nu} \rangle$, is distinct from the bulk stress tensor, $T_{ab}$, arising from the bulk Maxwell field. Explicitly, we find 
\begin{subequations}
    \begin{align}
        \langle T_{tt} \rangle &= - \frac{\nu \,\alpha_1(x)}{2 \ell_3},\quad \quad
        \langle T_{xx} \rangle = -\frac{\nu\left(\alpha_1(x)+3\,\alpha_3(x)\right)} {\ell_3 (2-x^2)(1-x^2)^2}, \quad\quad
        \langle T_{\phi\phi} \rangle =  \frac{\nu\left(3\,\alpha_3(x)-\alpha_1(x)\right)}{4 \ell_3 (1-x^2)^2}, \\
        \langle T_{tx} \rangle &= -\frac{3 \nu}{4\ell_3}\left(S_2'(x)+ \frac{4\,\alpha_4(x)}{(2-x^2)(1-x^2)}\right) 
        = -\frac{\nu\left(\mu \,g(x) \,C_1 +3 C_2\right)}{ \ell_3 \sqrt{2-x^2}}
    \end{align}
and
\begin{align}
    \langle J_t \rangle &= \frac{\nu\,\beta_6(x)}{\ell_3}, \quad\quad
    \langle J_x \rangle = \frac{\nu}{\ell_3} \left(\frac{\beta_7(x)}{\left(1-x^2\right)} + \mu\,g'(x) \right) 
    = \frac{\nu\,C_1}{\ell_3\sqrt{2-x^2}},
\end{align}
\end{subequations}
with $\nu = \ell_4^2/(4\pi G_4)$, and where we have made use of (\ref{eqn:beta}) and (\ref{eqn:alpha}) to simplify $A^{(1)}_x$ and $g^{(3)}_{tx}$. Figure \ref{fig:Tmunu} gives plots for the non-zero components of the holographic stress tensor for various values of the magnitude of the chemical potential, $\mu$.

One can explicitly check that the current is conserved:
\begin{equation}
    \mathrm{D}_\mu \langle J^\mu \rangle = 0,
\end{equation}
where $\mathrm{D}_\mu$ is the covariant derivative associated to the boundary geometry, and moreover the stress tensor is traceless and satisfies a Ward relation:
\begin{equation}
    \langle T_\mu{}^\mu \rangle = 0, \quad \mathrm{D}^\mu \langle T_{\mu\nu} \rangle - F^{(0)}_{\mu\nu}\langle J^\mu \rangle= 0,
\end{equation}
where $F^{0}$ is the field strength tensor arising from the source, (\ref{eqn:source}).
\section{The Ansatz in the DeTurck Gauge}
We also obtained the solutions via the DeTurck method, in which one adds additional terms to the equations of motion in order to make them elliptic. One solves the \textit{Einstein-DeTurck equations} given by
\begin{subequations}
    \begin{align}\label{eq:EDT}
    0 &= E^{H}_{ab} := R_{ab}+\frac{3}{\ell^2}g_{ab} - 2T_{ab} - \nabla_{(a}\xi_{b)} \,, \\
    0 &= \nabla^a F_{ab} - \nabla_b \zeta 
\end{align}
with
\begin{align}
    \xi^a &= g^{cd}\left[\Gamma^a_{cd}(g)-\Gamma^a_{cd}(\Bar{g})\right] \\
    \zeta &= g^{cd} \nabla_c \left(A_d - \Bar{A}_d\right)
\end{align}
\end{subequations}
where $\Gamma^a_{cd}(\mathfrak{g})$ is the Christoffel connection associated to a metric $\mathfrak{g}$, and $\Bar{g}_{ab}$ and $\Bar{A}_a$ are, respectively, a reference metric and a reference vector potential which we are free to choose. In order for solutions to the Einstein-DeTurck equations to also satisfy the Einstein-Maxwell equations, we need $\xi$ and $\zeta$ to vanish on the solutions --- otherwise the solutions are known as \textit{Ricci solitons}. In our case, in which there is a flowing geometry, there is, \textit{a priori}, no guarantee that the solutions we obtain via the DeTurck method will not be Ricci solitons. Therefore we must check after the fact that $\xi$ and $\zeta$ are small and tending to zero in the continuum limit.

The \textit{Ansatz} for the holographic batteries in DeTurck gauge is given by
\begin{subequations}\label{eq:dT_Ansatz}
\begin{align}
    \diff s^2 = \frac{\ell_4^2}{y^2} &\left[-p_1 \diff v^2 - 2 p_2 \diff v \diff y + p_5 \diff y^2 + \frac{4\,p_3}{2-x^2}\bigg(\frac{\diff x}{1-x^2} +p_6 \diff y + p_7 \diff v\bigg)^2 + \frac{p_4 \diff \phi^2}{\left(1-x^2\right)^2}\right]
\end{align}
and
\begin{equation}
    A = \ell_4 \left(p_8 \diff v + p_9 \diff y + \frac{p_{10}}{1-x^2} \diff x\right),
\end{equation}
\end{subequations}
where the unknown $p_i$ functions depend on $x$ and $y$. The reference metric and vector potential are obtained by setting
\begin{eqnarray}\label{reference}
    p_i(x,y) = \begin{cases}
        1-y^2 \quad\quad & \text{for}\; i=1 \\
        1 \quad &\text{for}\; 2\leq i \leq 4\\
        \mu \, g(x)\;&\text{for}\; i=8\\
        0 \quad &\text{otherwise}
    \end{cases}
\end{eqnarray}
The equations are again solved as a boundary value problem. Dirichlet boundary conditions are set at the conformal boundary, $y=0$, and the two asymptotic regions where the horizon meets the boundary at $x = \pm 1 $ simply by setting that the metric and vector potential are equal to the reference metric and reference vector potential, \textit{i.e.} $p_i$ is given by (\ref{reference}) at these boundaries. Similarly to in Bondi-Sachs gauge, these conditions enforce that the boundary metric is conformal to (\ref{eqn:bndy_metric}), the vector potential at the conformal boundary is equal to (\ref{eqn:source}) up to a factor of $\ell_4$, and that the bulk horizon approaches the geometry of a hyperbolic horizon near the boundary.

We do not set boundary conditions at the $y=1$ boundary. Instead, following the work of \cite{Figueras:2012rb}, we solve the equations of motion to a larger value of $y=y_{max}>1$, ensuring that the horizon of the obtained solution lies within the integration domain. This method implicitly enforces the condition that the metric is regular at the horizon, due to the fact that using pseudo-spectral method yields regular solutions throughout the interior of the integration domain.

Once the solutions are obtained, one can again extract the holographic stress tensor and conserved current via a process of holographic renormalization, though in DeTurck gauge one must contend with non-analyticities in the expansion. 

In this gauge the horizon is not necessarily the $y=1$ slice. One can find the position of the bulk horizon by assuming it takes the form $y=P(x)$ such that $P(1) = P(-1)=1$, and then the fact that the horizon must be a null hypersurface provides a first order ODE for $P$ which can be solved numerically. We find that the horizon ``bends inwards" in the coordinate domain so that $P(x) \leq 1 $ for all $x \in [-1,1]$, and so we are free to take $y_{max}$ as close to 1 as we like.

\begin{figure}[t!]
    \centering \includegraphics[width =0.75\textwidth]{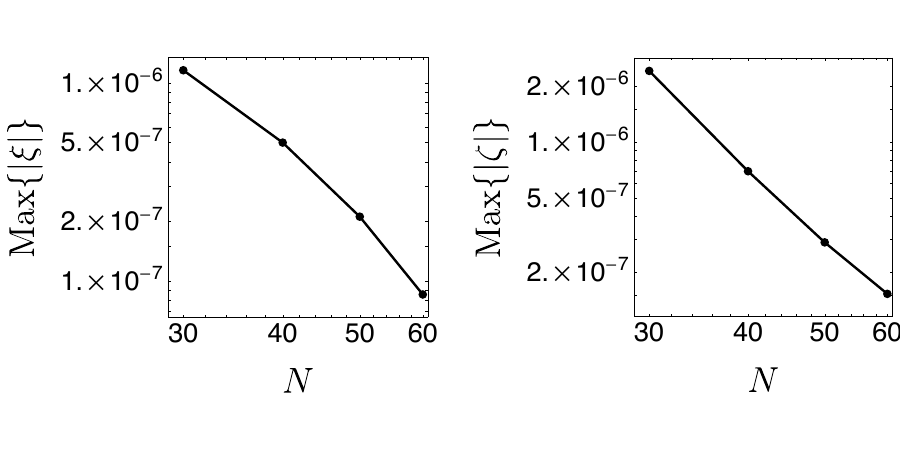}
    \caption{The maximum norm of the DeTurck vector on the left and of the DeTurck potential $\zeta$ on the right for a holographic battery with $\mu = 1$ in DeTurck gauge. The solutions were found numerically on a discrete lattice of size $2N \times N$ for different values of $N$, plotted in a log-log plot.}
    \label{fig:dT_convergence}
\end{figure}

We can track the DeTurck vector $\xi^a$ and $\zeta$ in order to study the convergence properties of the numerics in DeTurck gauge. We plot such convergence tests in Fig. \ref{fig:dT_convergence} for the holographic battery with $\mu = 1$. Once again the plots suggest power law convergence. Note that since the batteries are not static solutions, one may be concerned that the DeTurck vector could be a non-zero null vector. However, we also checked explicitly that each component of the DeTurck vector tend towards zero confirming that we really do have convergence to a solution to the Einstein-Maxwell equations.

Having found the solutions in two different gauges strongly supports that they are valid solutions to the Einstein-Maxwell equations, and we find that in the region of the parameter space in which we found solutions in both gauges all physical quantities matched. Though the use of Bondi-Sachs gauge is numerically less expensive and allows for the easier extraction of physical quantities, we found that the DeTurck gauge allowed us to obtain solutions with larger values of $\mu$ than was possible in Bondi-Sachs gauge.

\section{Conductance for other profiles}
\begin{figure}[t!]
    \centering
    \includegraphics[width=0.75\textwidth]{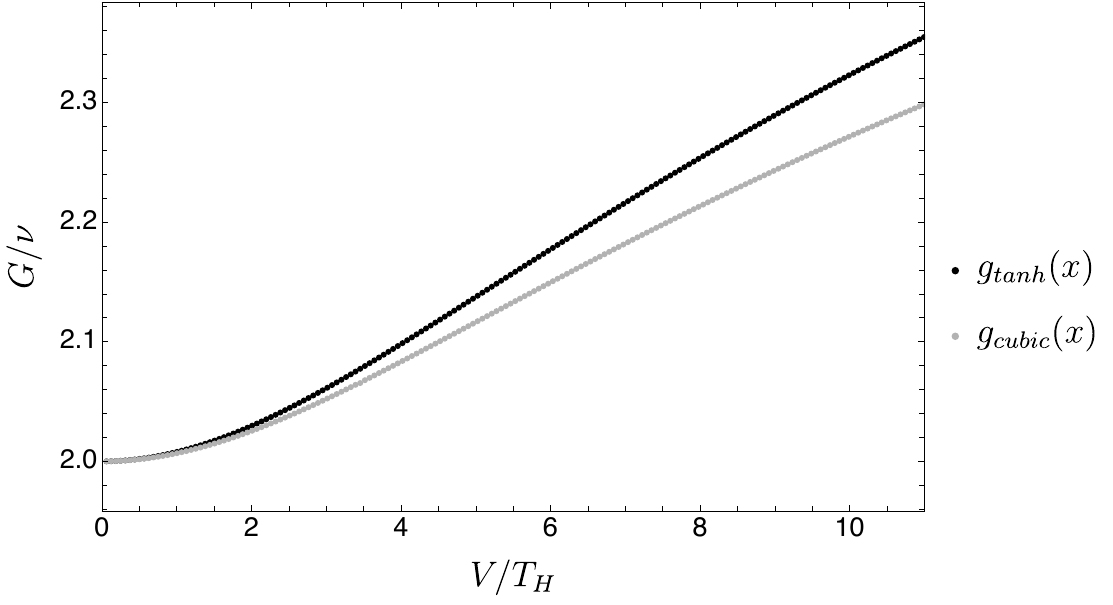}
    \caption{The conductance against the potential difference for holographic batteries with different profiles used for the source chemical potential. For the dots in black we used the tanh profile, whilst the dots in grey correspond to batteries defined with the cubic profile.}
    \vspace{-10pt}
    \label{fig:tanh_conductivity}
\end{figure}

We also computed holographic batteries for other choices of the chemical potential profile, $g(x)$. In each case, we fixed $g(1)=1=-g(-1)$ and $g'(1)=0=g'(-1)$. 

For example, we considered two other profiles, named the \textit{tanh profile} and the \textit{cubic profile} and given, respectively, by
\begin{align}
    g_{tanh}(x) := \tanh \left(\frac{x\sqrt{2-x^2}}{1-x^2} \right) \quad\quad \text{and} \quad\quad
    g_{cubic}(x) := \inv{2}\left(3x - x^3\right)
\end{align}
In Fig. \ref{fig:tanh_conductivity} we plot the conductance of the tanh and cubic profiles in black and grey, respectively.

Just as with the sine profile, we find that in each case $G/\nu \to 2$ as $V/T_H \to 0$; we provide a proof in the next section that this is a universal feature of all choices of profile, $g(x)$. Moreover, qualitatively, the shape of each of the conductance curves is very similar, with the conductance increasing with voltage. In each case the derivative of the conductance with respect to $V/T_H$ reaches a maximum, with the gradient of the curve decreasing after this point. The position of this turning point of the derivative depends upon the choice of profile.

\section{A proof of the linear behaviour}\label{sec:linear}
In Figures \ref{fig:conductance} and \ref{fig:tanh_conductivity} we see that the ratio $G / \nu$ tends to two in the limit $V / T_H \to 0$. In this section we provide a proof that this limiting behaviour occurs universally for any choice of the profile, $g(x)$, which is odd in $x$.

We begin by taking our \textit{Ansatz} (\ref{eq:ansatz}) with
\begin{subequations}
\begin{equation}
q_i=1\,\qquad \text{for}\qquad i \in \{1,2,3,5\}
\end{equation}
with
\begin{equation}
q_4=0\,,\qquad q_6=\delta \mu\,\hat{q}_6(x,y)\qquad\text{and}\qquad q_7=\delta \mu\,\hat{q}_7(x,y)
\end{equation}
\end{subequations}%
where $\delta \mu$ is taken to be arbitrarily small. To linear order in $\delta \mu$, the equations take the following simple form:
\begin{subequations}
\label{eqs:system}
\begin{equation}
4 (1-y^2) \frac{\partial ^2\hat{q}_6(x,y)}{\partial y^2}+(1-x^2)^2 \sqrt{2-x^2} \frac{\partial }{\partial x}\left(\sqrt{2-x^2} \frac{\partial \hat{q}_6(x,y)}{\partial x}\right)=0\,,
\end{equation}
\begin{equation}
\frac{\partial \hat{q}_7(x,y)}{\partial y}+\frac{1-x^2}{1-y^2}\left[\frac{A_2}{\sqrt{2-x^2}}+\frac{\partial \hat{q}_6(x,y)}{\partial x}\right]=0,
\end{equation}
\end{subequations}%
where $A_2$ is a constant to be determined in what follows. We now change to a new set of coordinates with which it is easier to explain our results. Let
\begin{equation}
X=x\sqrt{2-x^2}\,.
\end{equation}
In terms of $X$, the equations (\ref{eqs:system}) read
\begin{subequations}
\label{eqs:systemX}
\begin{equation}
 (1-y^2) \frac{\partial ^2\hat{q}_6(X,y)}{\partial y^2}+(1-X^2)^{3/2} \frac{\partial }{\partial x}\left(\sqrt{1-X^2} \frac{\partial \hat{q}_6(X,y)}{\partial X}\right)=0\,,
 \label{eq:systemXa}
\end{equation}
\begin{equation}\label{eq:dyhatq7}
\frac{\partial \hat{q}_7(X,y)}{\partial y}+\frac{1}{1-y^2}\frac{\sqrt{1-X^2}}{\sqrt{1+\sqrt{1-X^2}}}\left[A_2+2\sqrt{1-X^2}\frac{\partial \hat{q}_6(X,y)}{\partial X}\right]=0\,.
\end{equation}
\end{subequations}%
The general solution to Eq.~(\ref{eq:systemXa}) can be written as
\begin{subequations}
\label{eqs:hatqs}
\begin{equation}
\hat{q}_6(X,y)=(A_0+A_1\,y)\arcsin X+(1-X^2)^{1/4}\int_{-\infty}^{+\infty}{\rm d}q\,b(q)\,S_q(X)K_q(y)\,
\label{eq:hatq6}
\end{equation}
with
\begin{equation}
\frac{\partial }{\partial X}\left[(1-X^2) \frac{\partial S_q(X)}{\partial X}\right]+\left(\frac{q^2}{1-X^2}-\frac{1}{4}\right) S_q(X)=0
\label{eq:S}
\end{equation}
and
\begin{equation}
(1-y^2) \frac{\partial ^2K_q(y)}{\partial y^2}-\frac{1}{4} \left(1+4 q^2\right) K_q(y)=0\,.
\label{eq:K}
\end{equation}
Meanwhile, for $\hat{q}_7$, we find that
\begin{multline}
\hat{q}_7(X,y)=Z(X)+\frac{\sqrt{1-X^2}}{2\sqrt{1+\sqrt{1-X^2}}}\left[(2A_0+2 A_1+A_2) \log(1-y)-(2A_0-2 A_1+A_2) \log(1+y)\right]
\\
+\frac{4(1-X^2)^{1/4}}{\sqrt{1+\sqrt{1-X^2}}}\int_{-\infty}^{+\infty} {\rm d}q\,\frac{b(q)}{1+4 q^2}\left[X S_q(X)-2 (1-X^2) \frac{\partial S_q(X)}{\partial X}\right] \frac{\partial K_q(y)}{\partial y}
\label{eq:hatq7}
\end{multline}
\end{subequations}%
where $Z(X)$ is an integration function, and $A_0$, $A_1$ and $A_2$ are constants to be determined in what follows.

General complex solutions to Eq.~(\ref{eq:S}) can be written as
\begin{equation}
\tilde{S}_q(X)=z_1(q) P_{-1/2}^{i q}(X)+z_2(q) Q_{-1/2}^{i q}(X)
\end{equation}
where $P_{\mu}^{\nu}(X)$ are associated Legendre functions of the first kind of order $\mu$ and degree $\nu$, $Q_{\mu}^{\nu}(X)$ are associated Legendre functions of the second kind of order $\mu$ and degree $\nu$ and $z_1(q)$ and $z_2(q)$ are arbitrary integration constants.

On the other hand, general complex solutions to Eq.~(\ref{eq:K}) can be written as
\begin{equation}
\tilde{K}_q(y)=p_1(q)\, _2F_1\left(-\frac{1}{4}-\frac{i q}{2},-\frac{1}{4}+\frac{i q}{2};\frac{1}{2};y^2\right) +p_2(q)\,y \, _2F_1\left(\frac{1}{4}-\frac{i q}{2},\frac{1}{4}+\frac{i q}{2};\frac{3}{2};y^2\right)
\end{equation}
where $p_1(q)$ and $p_2(q)$ are arbitrary integration constants, while $\, _2F_1(a,b;c;z)$ is a Gauss hypergeometric function. Regularity at $y=1$ (the future event horizon) demands that
\begin{equation}
p_2(q)=-\frac{2 \Gamma \left(\frac{1}{4}-\frac{i q}{2}\right) \Gamma \left(\frac{i q}{2}+\frac{1}{4}\right)}{\Gamma \left(-\frac{i q}{2}-\frac{1}{4}\right) \Gamma \left(\frac{i q}{2}-\frac{1}{4}\right)}p_1(q)
\end{equation}
so that
\begin{equation}
\tilde{K}_q(y)=\frac{\Gamma \left(\frac{5}{4}-\frac{i q}{2}\right) \Gamma \left(\frac{5}{4}+\frac{i q}{2}\right)}{\sqrt{\pi }} (1-y^2) \, _2F_1\left(\frac{3}{4}-\frac{i q}{2},\frac{3}{4}+\frac{i
   q}{2};2;1-y^2\right)\,
\end{equation}
and in particular we have chosen $p_1(q)=1$ above so that $\tilde{K}_q(0)=1$. Note that we necessarily have $\tilde{K}_q(1)=0$. Note that we want $q_6$ and $q_7$ to be \emph{real} functions, and the symmetries of the hypergeometric function are such that
\begin{equation}
K_q(y)=\frac{1}{2}(\tilde{K}_q(y)+\tilde{K}_{-q}(y))=\tilde{K}_q(y)\,.
\end{equation}

Next we look at the boundary conditions at $X=\pm1$ (i.e. $x=\pm1$). Here we want $\hat{q}_6(\pm 1,y)=\pm1$, so that $q_6(\pm1,y)=\pm \delta \mu$. It is a simple exercise to show that near $X=\pm1$ we have $S_1(X)\sim(1\pm X)^{\frac{i q}{2}}+\gamma\,(1\pm X)^{\frac{-i q}{2}}$, with $\gamma$ a constant. This means that the last term in Eq.~(\ref{eq:hatq6}) does not contribute near $X=\pm1$, so long as $b(q)$ does not grow very large as $q\to+\infty$. As such, our boundary conditions demand that $A_1=0$ in Eqs.~(\ref{eqs:hatqs}).

Note that regularity at the future event horizon, located at $y=1$, also demands that $A_2=-2A_0$, so that no $\log(1-y)$ term appears on the right hand side of Eq.~(\ref{eq:hatq7}). Moreover, note that by comparing Eq.~\eqref{eq:dyhatq7} with Eq.~\eqref{eq:dyq7} in the main text, one finds that $C_1 = -A_2 \delta \mu$.
Therefore, at this stage we only need to connect $A_0$ with $\mu$ appearing the main text.

Thus far we have imposed all boundary conditions, except those at the conformal boundary. Firstly, we choose $Z(X)$ in Eq.~(\ref{eq:hatq7}) so that $\hat{q}_7(X,0)=0$, that is to say
\begin{equation}
Z(X)=\frac{8(1-X^2)^{1/4}}{\sqrt{1+\sqrt{1-X^2}}}\int_{-\infty}^{+\infty} {\rm d}q\,\frac{b(q)}{1+4 q^2}\left[X S_q(X)-2 (1-X^2) \frac{\partial S_q(X)}{\partial X}\right] \frac{\Gamma \left(\frac{1}{4}-\frac{i q}{2}\right) \Gamma \left(\frac{1}{4}+\frac{i q}{2}\right)}{\Gamma \left(-\frac{1}{4}-\frac{i q}{2}\right) \Gamma \left(-\frac{1}{4}+\frac{i q}{2}\right)}\,.
\end{equation}

We are thus left to determine $b(q)$ for a given profile. Let the profile be given as in the main text by $q_6(X,0)=\delta \mu g(X)$ with $g(X)$ and odd function of $X$ satisfying $g(1)=1$. This means we want to choose the $S_q(X)$ to be real and odd. Demanding that $S_q(X)$ is odd, implies
\begin{equation}
z_2(q)=-\frac{1}{\pi }\left(2 i+\frac{4}{i+e^{\pi  q}}\right)z_1(q)\,.
\end{equation}
while demanding that $S_q(X)$ is real, imposes
\begin{equation}
S_q(X)=\frac{1}{2}\left(\tilde{S}_q(X)+\overline{\tilde{S}_q(X)}\right)\,
\end{equation}
where a bar denotes complex conjugation.

Using known properties of the associated Legendre functions and the Dirac delta distributions, one can prove that
\begin{subequations}
\begin{equation}
\int_{-1}^{1}\mathrm{d}X\,\frac{S_q(X)S_p(X)}{1-X^2}=\eta(p)[\delta (p-q)+\delta (p+q)]
\end{equation}
with
\begin{equation}
\eta(p)=\frac{1}{p \pi }\Bigg\{\frac{2 \pi }{\sinh p \pi}+\left[\frac{1}{\sinh p \pi}-i\right] \Gamma \left(\frac{1}{2}-i p\right)^2+\left[\frac{1}{\sinh p \pi}+i\right] \Gamma \left(\frac{1}{2}+i p\right)^2\Bigg\}\,,
\end{equation}
\end{subequations}
where $\delta(x)$ is a Dirac delta function.

The procedure to determine $b(q)$ (and thus the full bulk solution) from a given profile is now clear. First, from Eq.~(\ref{eq:hatq6}) with $A_1=0$, we demand
\begin{equation}
g(X)=A_0 \arcsin X+(1-X^2)^{1/4}\int_{-\infty}^{+\infty}{\rm d}q\,b(q)\,S_q(X)
\end{equation}
Note that $g(1)=1$, so that we must take $A_0=\frac{2}{\pi}$ and hence, after rearranging we have
\begin{equation}
\frac{1}{(1-X^2)^{1/4}}\left[g(X)-\frac{2}{\pi}\arcsin X\right]=\int_{-\infty}^{+\infty}{\rm d}q\,b(q)\,S_q(X)\nonumber
\end{equation}
and thus
\begin{align}
\int_{-1}^{1}{\rm d}X\frac{S_p(X)}{(1-X^2)^{5/4}}\left[g(X)-\frac{2}{\pi}\arcsin X\right]& =\int_{-1}^{1}{\rm d}X\int_{-\infty}^{+\infty}{\rm d}q\,b(q)\,\frac{S_p(X)\,S_q(X)}{1-X^2}\nonumber
\\
&=\int_{-\infty}^{+\infty}{\rm d}q\,b(q)\,\int_{-1}^{1}{\rm d}X\,\frac{S_p(X)\,S_q(X)}{1-X^2}\nonumber
\\
&=\eta(p)\left[b(p)+b(-p)\right]\,.
\end{align}
so that
\begin{equation}
b(p)+b(-p)=\frac{1}{\eta(p)}\int_{-1}^{1}{\rm d}X\frac{S_p(X)}{(1-X^2)^{5/4}}\left[g(x)-\frac{2}{\pi}\arcsin X\right]\,.
\end{equation}
Finally, because $S_p(X)=S_{-p}(X)$ and $K_p(y)=K_{-p}(y)$, it follows that
\begin{equation}
\hat{q}_6(X,y)=\frac{2}{\pi}\arcsin X+(1-X^2)^{1/4}\int_{0}^{+\infty}{\rm d}q\,[b(q)+b(-q)]\,S_q(X)K_q(y)\,
\end{equation}
thus completing the bulk solution for a generic boundary profile $g(X)$.

Overall, we have learned that $A_0=2/\pi$, and hence $A_2=-4/\pi$, which in turn implies that, in the notation of the main text, $C_1 = 4\delta\mu/\pi$. Finally applying the definition of the conductance, \eqref{eqn:conductance}, for this small amplitude, $\delta \mu$, one finds that $G/\nu = 2$, irrespective of the choice of odd profile, $g(x)$.

\end{document}